\documentclass[twoside]{article}  
\usepackage{fleqn,espcrc2}
\usepackage{graphicx}
\usepackage{epsf,psfrag}
\usepackage[figuresright]{rotating}

\newcommand{\bda}{\begin{displaymath}\begin{array}{rl}}
\newcommand{\eda}{\end{array}\end{displaymath}}
\newcommand{\be}{\begin{equation}}
\newcommand{\ee}{\end{equation}}
\newcommand{\bea}{\begin{eqnarray}}
\newcommand{\eea}{\end{eqnarray}}
\newcommand{\bdm}{\begin{displaymath}}
\newcommand{\edm}{\end{displaymath}}
\newcommand{\no}{\nonumber \\}

\newcommand{\ubar}{\overline{\rule[0.42em]{0.4em}{0em}}\hspace{-0.5em}u}
\newcommand{\dbar}{\,\overline{\rule[0.7em]{0.4em}{0em}}\hspace{-0.6em}d}
\newcommand{\sbar}{\overline{\rule[0.45em]{0.4em}{0em}}\hspace{-0.5em}s}

\newcommand{\Kbar}{\,\overline{\rule[0.75em]{0.7em}{0em}}\hspace{-0.85em}K}

\newcommand{\QCD}{{\mbox{\scriptsize Q\hspace{-0.1em}CD}}}

\newcommand{\R}{{\mbox{\tiny R}}}
\renewcommand{\L}{{\mbox{\tiny L}}}
\newcommand{\al}{&\!\!\!\!}
\newcommand{\fs}{\;  .}
\newcommand{\co}{\;  ,}

\newcommand{\ind}{\scriptscriptstyle}

\title{Non-lattice determinations of the light quark masses}

\author{H. Leutwyler\address[MCSD]{Institute for Theoretical
        Physics, University of Bern\\Sidlerstr. 5, CH-3012 Bern, Switzerland}}

\begin{document}

\begin{abstract}\noindent{\it\large Talk given at Lattice 2000, 
Bangalore, India, Aug. 2000.}

\vspace{0.7em}\noindent 
The sum rule determinations of $m_s$ and the chiral 
perturbation theory results for the ratios $m_u/m_d$ and 
$m_s/m_d$ are reviewed and a method for the extrapolation of
lattice data to the physical values of $m_u$ and $m_d$ is outlined.
\vspace{1pc}
\end{abstract}

\maketitle

\section{INTRODUCTION}
The first crude estimates \cite{Phys Rep} for the magnitude of the three
lightest quark masses appeared 25 years ago (values in MeV):
\begin{displaymath}
\begin{array}{llll}
m_u\, \simeq \,4\co & m_d \,\simeq\, 6\co
&  m_s\, \simeq
135\co &\cite{GL75} \\
m_u\, \simeq 4.2\co & m_d \,\simeq 7.5\co
& m_s \,\simeq
150\fs &\cite{Weinberg1977}
\end{array}\end{displaymath}
Many papers dealing with the pattern of quark masses have been published since
then.\footnote{For an extensive list of refs., see \cite{Manohar}. Like
  most of these, the present brief report is based on the hypothesis that the
  quark condensate represents 
the leading order parameter of the spontaneously broken symmetry.
The underlying physics issues are discussed in \cite{Descotes Stern} and the
refs.~therein.} 

One of the reasons for being interested in an accurate determination of the
mass values is that these are not understood at all -- we need to know the
numbers to test ideas that might lead to an understanding of the
pattern, such as the relations with the lepton masses that emerge from 
attempts at
unifying the electroweak and strong forces. Another reason is that the
Standard Model must describe the low energy properties of the various
particles to an amazing degree of accuracy. At low energies, the weak
interactions are 
frozen and the neutrini decouple -- the Standard Model reduces to the gauge
field theory of SU(3)$\times$U(1). In the framework of this effective theory,
the quark 
masses occur as para\-meters in the Lagrangian, together with the coup\-ling
constants $e,g$, the vacuum angle $\theta$ and the masses of the charged
leptons. This Lagrangian is supposed to describe the low energy structure of
cold matter to a very high degree of precision, provided the parameters
occurring therein are
accurately known. The size of the atoms, for instance, is determined by
$a_{\hspace{-0.05em}\ind B}=4\pi/(e^2m_e)$ and thus only involves parameters
for which this is the case. 

In the following, I discuss the three light quark masses in terms of
the magnitude of $m_s$ and of the ratios $m_u/m_d$ and $m_s/m_d$, which
characterize their relative size.

\section{MAGNITUDE OF {\large\boldmath $m_s$ \unboldmath}}
Apart from the lattice approach, the best determinations of the magnitude of
$m_s$ rely on QCD sum rules \cite{QCD SR1}. For a recent, comprehensive
review, I refer to \cite{CK}.
A detailed discussion of the method in application to the mass spectrum of
the quarks was given nearly 20 years ago \cite{Phys Rep}.
The result for the $\overline{\mbox{MS}}$ running mass at scale
$\mu=1\,\mbox{GeV}$ obtained at that time was
$m_s(1\,\mbox{GeV})=175\pm 55\,\mbox{MeV}$. 

The issue has been investigated in
considerable detail since then: Various versions of the sum rules for the
pseudoscalar and scalar correlation functions were studied, as well as
sum rules for two-point functions formed with baryonic currents.
More recently, an entirely independent sum rule determination of $m_s$ 
based on the strangeness content of the final state in $\tau$-decay
became possible -- in principle, this source of information is the cleanest
one, because the hadronic input does not involve theoretical models, but
can be taken from experiment. The analy\-sis of the $\tau$ decay data
described in \cite{Pich  Prades} leads to\footnote{Note the difference in
  scale: In recent times 
it has become customary to quote values at the scale $\mu=2\,\mbox{GeV}$,
because perturbation theory is under better control there. With the conversion
factor used in \cite{Pich Prades}, the old value mentioned above corresponds to
$m_s(2\,\mbox{GeV})=125\pm 40 \,\mbox{MeV}$.}   
$m_s(2\,\mbox{GeV})=114 \pm 23 \,\mbox{MeV}$. 
As discussed there and in \cite{Kambor Maltman}, 
this result supersedes the earlier one 
in \cite{ALEPH}, which is based on the same experiment and on the same 
theoretical framework.

The various sum rule results for $m_s$ are reviewed in 
\cite{CK,Narison}. In \cite{Narison}, the ``world average'' is quoted as
$m_s (2\,\mbox{GeV})=118.9\pm 12.2\,\mbox{MeV}$, corroborating the 
above numbers with a significantly smaller error bar. It is notoriously
difficult, however, to account for the systematic errors of
the various investigations \cite{ms sum rules} -- in my opinion,
the number given does not fully cover these.

The results based on the lattice approach, which were reported at this
conference, cluster around $m_s(2\,\mbox{GeV})\simeq 90 \,\mbox{MeV}$, 
that is at the lower edge of the range obtained on the basis of sum rules. 
I think that the results are
consistent with one another. In my opinion, this is
important, because it shows that 
the magnitude of $m_s$ is understood -- we should not be content with
numerical simulations of the theory, even if these will in the long run yield
the most accurate results. A problem would arise, however,  
if the gradual reduction of the systematic errors to be attached to 
the lattice results should turn out to push the result downwards -- 
that would be difficult to 
re\-con\-cile with the sum rules, in particular also
in view of the lower bounds derived from the positivity
of the spectral functions \cite{lower bound}. 

\section{PSEUDOSCALAR MASSES}
\label{pseudoscalar masses}
The best determinations of the
{\it relative} size of $m_u$, $m_d$ and $m_s$
rely on the fact that these masses
happen to be small, so that the properties of the theory
may be analyzed by treating the quark mass term in the Hamiltonian of QCD
as a perturbation. The Hamiltonian is split into two pieces:
\bdm H_\QCD=H_0+H_1\co\edm
where $H_0$ describes the three lightest quarks as massless and $H_1$ is the
corresponding mass term,
\bdm H_1=\!\int\!d^3\!x\{m_u\,\ubar u+m_d \,\dbar d+ m_s\, \sbar s\} \fs\edm
$H_0$ is invariant under
the group SU(3)$_\R\times$SU(3)$_\L$ of independent flavour rotations of
the right- and lefthanded quark fields. The symmetry is broken
spontaneously: The eigenstate of $H_0$ with the lowest
eigenvalue, $|0\rangle$, is invariant only under the subgroup
SU(3)$_{\mbox{\tiny V}}\!\subset\,$SU(3)$_\R\times$SU(3)$_\L$.
Accordingly, the spectrum of $H_0$ contains
eight Goldstone bosons, $\pi^\pm,\pi^0,K^\pm,K^0,\Kbar^0,\eta$. The remaining
levels form degenerate multiplets of SU(3)$_{\mbox{\tiny V}}$ of non-zero
mass.

The perturbation $H_1$ splits
the SU(3) multiplets, in particular also the Goldstone boson octet. To first
order, the squares of the meson masses are linear in $m_u,m_d,m_s$ and the
symmetry fixes the coefficients up to a 
constant:
\bea M_{\pi^+}^2\al=\al(m_u+m_d)B_0+O(m^2)\co\no
M_{K^+}^2\al=\al (m_u+m_s)B_0+O(m^2)\co\no
     M_{K^0}^2\al=\al (m_d+m_s)B_0+O(m^2)\fs\nonumber\eea
In the ratios $M_{\pi^+}^2\!:\!M_{K^+}^2\!:\!M_{K^0}^2$, the constant $B_0$
drops 
out. Using the Dashen theorem \cite{Dashen} to account for the e.m.~self 
energies (see section \ref{sec:Q}), these relations imply
\cite{Weinberg1977} \bea \label{i3}
\frac{m_u}{m_d} \al\simeq\al \frac{M^2_{K^+}-  M^2_{K^0} + M^2_{\pi^+}} 
{M^2_{K^0} - M^2_{K^+} + M^2_{\pi^+}}\simeq 0.55\co \no
\frac{m_s}{m_d} \al\simeq\al \frac{M^2_{K^0} + M^2_{K^+} - M^2_{\pi^+}}
{M^2_{K^0} - M^2_{K^+} + M^2_{\pi^+}}\simeq 20.1\fs
\eea

\section{SECOND ORDER MASS FORMULAE}
The leading order mass formulae are subject to corrections
arising from contributions which are of second or higher order in the
perturbation $H_1$. These can systematically be analyzed by means of the
effective Lagrangian method \cite{Weinberg Physica,GL SU(3)}.  
In this approach, the quark and gluon fields of QCD are replaced by a
set of pseudoscalar fields describing the degrees of freedom of the Goldstone
bosons $\pi, K, \eta$. The effective Lagrangian
only involves these fields and their derivatives, but contains an infinite
string of vertices. For the calculation of the pseudoscalar masses
to a given order in the perturbation $H_1$, however, only a finite subset
contributes. The term $\Delta_M$, which describes
the SU(3) corrections in
the ratio $M_K^2/M_\pi^2$,
\bdm
\label{i5}\frac{M_K^2}{M_\pi^2}=
\frac{\hat{m}+m_s}{m_u+m_d}\{1+\Delta_M+O(m^2)\}\co\edm
involves the two
effective
coupling constants $L_5$, $L_8$, which occur in the derivative expansion 
of the effective Lagrangian at first non-leading order \cite{GL SU(3)}:
\bdm
\label{DeltaM}\Delta_M=
\frac{8(M_K^2-M_\pi^2)}{F_\pi^2}\,(2L_8-L_5)+\chi\mbox{logs}\fs\edm
The term $\chi$logs stands for the logarithms characteristic of
chiral perturbation theory. They arise because the spectrum of the unperturbed
Hamiltonian $H_0$ contains massless particles -- the
perturbation $H_1$ generates infrared singularities.
The coupling constant $L_5$ also determines the SU(3) asymmetry in the decay
constants,
\bdm\label{DeltaF}\frac{F_K}{F_\pi}=1+\frac{4(M_K^2-M_\pi^2)}{F_\pi^2}\,L_5
+\chi\mbox{logs}\fs\edm
The first order SU(3) correction in the
mass ratio $(M_{K^0}^2-M_{K^+}^2)/(M_K^2-M_\pi^2)$ turns out to be 
the same as the one in
$M_K^2/M_\pi^2$ \cite{GL SU(3)}:
\bdm \frac{M_{K^0}^2-M_{K^+}^2}{M_K^2-M_\pi^2}=\frac{m_d-m_u}{m_s-\hat{m}}
\{1+\Delta_M +O(m^2)\}\fs\edm
Hence, the first order corrections drop out in the double ratio
\be\label{defQ}
Q^2 \equiv \frac{M^2_K}{M_\pi^2}\cdot \frac{ M^2_K - M^2_\pi}{M^2_{K^0} -
M^2_{K^+}}\fs \end{equation}
The observed values of the meson masses thus provide a tight constraint on 
one particular ratio of quark masses:
\be
Q^2 = \frac{m^2_s - \hat{m}^2}{m^2_d - m^2_u} \{ 1 + O (m^2) \}\fs
\end{equation}
The constraint may be visualized by
plotting the ratio
$m_s/m_d$ versus $m_u/m_d$ \cite{Kaplan Manohar}.
Dropping
the higher order contributions, the
resulting curve takes the form of an ellipse:
\be\label{ellipse}
\left ( \frac{m_u}{m_d} \right)^2 + \,\frac{1}{Q^2} \left ( \frac{m_s}{m_d}
\right)^2 = 1\co
\end{equation}
with $Q$ as major semi-axis (the term $\hat{m}^2/m_s^2$ has been discarded,
as it is
numerically very small).

\section{VALUE OF {\large\boldmath $Q$\unboldmath}}\label{sec:Q}
The meson masses occurring in the double ratio (\ref{defQ}) refer to pure QCD.
The Dashen theorem states that in the chiral limit, the electromagnetic
contributions to $M_{K^+}^2$ and to $M_{\pi^+}^2$ are the same, while the self
energies of $K^0$ and $\pi^0$ vanish.
Since the contribution to the
mass difference between $\pi^0$ and $\pi^+$ from $m_d\!-\!m_u$ is
negligibly small, the masses in pure QCD are approximately given by
\bea (M_{\pi^+}^2)^\QCD\al=\al
(M_{\pi^0}^2)^\QCD= M_{\pi^0}^2\co\no (M_{K^+}^2)^\QCD\al=\al
M_{K^+}^2-M_{\pi^+}^2+M_{\pi^0}^2\co\no(M_{K^0}^2)^\QCD\al=\al M_{K^0}^2\co
\nonumber\eea
where $M_{\pi^0},M_{\pi^+},M_{K^0},M_{K^+}$ are the
observed masses. Correcting
for the electromagnetic self energies in this way, relation (\ref{defQ}) 
yields $Q=24.2$. 
For this value of the semi-axis, the ellipse passes through the point specified
by Weinberg's mass ratios, eq.~(\ref{i3}).
The Dashen theorem is subject to corrections from higher order terms in the
chiral expansion, which have been analyzed, but are not fully
understood. Since the $K^0-K^+$ mass difference is dominated by the
contribution from $m_d-m_u$, the uncertainties in the e.m.~part do not very
strongly affect the value of $Q$, but they do show up at the 
5\% level. For a detailed discussion, I refer to \cite{Light quark masses}.

The isospin violating decay $\eta \rightarrow 3\pi$ allows one to measure the
semi-axis in an entirely independent manner \cite{GL eta}. The transition
amplitude is much less sensitive to the
uncertainties associated with the electromagnetic interaction than the
$K^0\!-\!K^+$ mass difference: The e.m.~contribution is
suppressed by chiral symmetry and is negligibly small \cite{Baur Kambor
  Wyler}. The transition amplitude thus represents a sensitive probe of the
symmetry breaking generated by $m_d-m_u$. In fact, the decay rate is 
proportional to $Q^{-4}$, with a factor that can be calculated up to
corrections of second order in the quark masses. The result for the semi-axis
is consistent with the information obtained from $K^0-K^+$ but is more
accurate \cite{Kambor Wiesendanger Wyler,AL}: 
\bea Q=22.7\pm 0.8\fs\eea

\section{KM--AMBIGUITY}\label{KM}
Chiral perturbation theory thus fixes the isospin breaking parameter $m_u/m_d$
in terms of the SU(3) breaking parameter $m_s/m_d$, to within small
uncertainties. The ratios themselves, 
that is the position on the ellipse, are a more subtle issue. Kaplan and
Manohar \cite{Kaplan Manohar} pointed out that the
corrections to the lowest order result, eq.~(\ref{i3}),
cannot be determined on purely phenomenological grounds.
They
argued that these corrections might be large and that the $u$-quark might
actually be massless. A couple of years ago, this possibility was widely
discussed in the literature \cite{Banks Nir Seiberg}, in view of
the strong CP problem. 

The reason why phenomenology alone does not allow us to determine the
two individual ratios beyond leading order is the following.
The matrix \bdm m' = \alpha_1 m + \alpha_2 (m^+)^{-1} \det m
\edm
transforms in the same manner as $m$. Symmetry does therefore not distinguish
$m'$ from $m$. 
For a real, diagonal mass matrix, the transformation law for $m_u$, for
instance, reads
\bea\label{KM1}
m_u' = \alpha_1 m_u + \alpha_2\, m_d\, m_s  \fs\eea
Since the effective
theory exclusively exploits the symmetry properties of QCD, the above
transformation of the quark mass matrix does not change the form of the
effective Lagrangian -- the transformation may be absorbed in a suitable
change of the effective coupling constants \cite{Kaplan Manohar}. This implies,
however, that the expressions for the masses of the pseudoscalars, 
for the scattering amplitudes or for the matrix elements of the vector and
axial
currents, which follow from this Lagrangian, are invariant under the operation
$m\!\rightarrow
\!m'$. Conversely, the experimental information on these observables
does
not distinguish $m_u$ from $m_u'$. Indeed, one readily checks that the
transformation $m\!\rightarrow\!m'$ maps the ellipse onto itself (up to terms
of order
$(m_u - m_d)^2/m_s^2$,
which were neglected).
Since the position on the ellipse does not remain invariant, it cannot be
extracted from these observables within chiral perturbation theory.

One is not dealing with a hidden symmetry of QCD here -- this
theory is not invariant under $m\rightarrow m'$. 
Some authors have made the claim that there is no reason for the quark masses 
entering the Lagrangian of the
effective theory to be the same as the running masses of QCD. 
This claim is incorrect, as can explicitly be demonstrated with
the following simple example. The
Ward identity for the axial current implies that
the vacuum-to-pion matrix element of the pseudoscalar density is given
by\footnote{The relation involves the matrix element of a 
pseudoscalar operator -- these are not KM-invariant, even at the level of the
effective theory.} 
\bdm\langle 0|\bar{d}\,i\gamma_5 u|\pi^+\rangle=\sqrt{2}\,F_{\pi}
M_{\pi^+}^2/(m_u+m_d)\fs\edm
The relation is exact, except for electroweak corrections (in the
$\overline{\mbox{MS}}$ scheme, both the matrix ele\-ment and the quark masses 
depend on the running scale, but the relation holds at any scale). 
In the framework of the effective theory, the two sides can
separately be calculated in terms of the effective coupling
constants and the quark masses occurring in the effective 
Lagrangian. The relation is obeyed, order by order in the chiral
perturbation series. What appears on the right, however,
are the quark masses of the effective Lagrangian. The
effective theory thus reproduces the matrix element 
$\langle 0|\bar{d}\,i\gamma_5 u|\pi^+\rangle$ 
if and only if the quark masses in the effective Lagrangian are identified 
with the running masses of QCD.

\section{THE RATIOS {\large\boldmath
    $m_u\!\!:\!m_d$\unboldmath} AND {\large\boldmath
    $m_s\!\!:\!m_d$\unboldmath}}\label{ratios}

The KM-ambiguity is of phenomenological nature: Unfortunately, an 
experimental probe sensitive to the 
scalar or pseudoscalar currents does not exist -- the electromagnetic
and weak interactions happen to probe the low energy structure of QCD 
exclusively through
vector and axial currents. This means that theoretical input 
is needed to determine the size of the two individual mass ratios
beyond leading order. 

One approach relies on the expansion in $1/N_c$. Since the KM-tranformation
(\ref{KM1}) violates the Okubo-Iizuka-Zweig rule, the parameter 
$\alpha_2$ is suppressed in the large $N_c$ limit. In order to cover 
this limit, the effective theory must be 
extended, including the degrees of freedom of the $\eta'$ among the 
dynamical variables, because this particle becomes massless if 
the number of colours is sent to infinity and the quark
masses are turned off. In fact, 
within that framework, the KM-ambiguity disappears 
altogether: The transformation $m'\rightarrow m$ preserves the large 
$N_c$ counting rules only if $\alpha_2$ vanishes to all orders in $1/N_c$ 
\cite{Kaiser Leutwyler}. An evaluation of the mass ratio $m_s/\hat{m}$ based
on the large $N_c$ analysis of the transitions 
$\eta'\rightarrow\gamma\gamma$ and $\eta\rightarrow\gamma\gamma$ is described
in \cite{Montpellier}.  

Another method is based on the hypothesis that the lowest resonances  
dominate the low energy behaviour of the various
Green functions -- once the poles and cuts due to the exchange of the
Goldstone bosons are accounted for. More precisely, one first shows that
the effective coupling constants can be represented in terms of convergent
integrals over suitable spectral functions and then assumes that these 
sum rules are approximately saturated by resonances \cite{GL
  1984,Ecker,Leutwyler1990,ABT}.  
 
Both of these methods lead to the conclusion that the corrections to the
lowest order mass formulae are small:
\bea \frac{m_u}{m_d}=0.553\pm0.043\co\;\;\;
     \frac{m_s}{m_d}=18.9\pm0.8\fs\eea
For a detailed discussion of this result, I refer to 
\cite{Light quark masses}. The values for the ratio $(m_u+m_d)/m_s$ 
obtained by means of sum rules 
\cite{Narison} as well as the lattice results 
presented at this conference are in remarkably good
agreement with these numbers. 

\section{EXTRAPOLATION IN {\large\boldmath $m_u\!\!$ \unboldmath} AND 
{\large\boldmath $m_d$ \unboldmath}}\label{extrapolation}
The remainder of this report concerns 
the interface between the lattice approach and the effective theory.
As pointed out long ago \cite{finite volume}, chiral perturbation theory
allows one to predict the dependence of the various correlation functions 
on the quark masses,  as well as on the temperature and on the volume of 
the box used. In the following, I only
discuss the predictions for the dependence on the quark masses. 

By now, dynamical quarks with a mass of the order of 
the physical value of $m_s$ are within reach, but it is notoriously 
difficult to equip the two lightest quarks with their proper masses. 
Suppose that we keep $m_s$ fixed at the physical value and 
set $m_u=m_d=\hat{m}$, 
but vary the value of $\hat{m}$ in the range $0< \hat{m}<\frac{1}{2}\,m_s$ (at
the upper end of that range, the pion mass is about 500 MeV). 

The expansions of $M_\pi$ and $F_\pi$ in powers of $\hat{m}$ are 
known to next-to-next-to leading order, from a two loop
analysis \cite{Buergi,BCEGS,BCE} 
of the effective field theory based on 
SU(2)$_\R\times$SU(2)$_\L$:
\bea \frac{M_\pi^2}{M^2}\al=\al 1-\mbox{$\frac{1}{2}$}\,x\,\bar{\ell}_3+
\mbox{$\frac{17}{8}$}\,x^2 \bar{\ell}_{\ind M}^{\;2}+
x^2 k_{\ind M}+O(x^3)\co\no
\frac{F_\pi}{F}\al=\al1+x\,\bar{\ell}_4-\mbox{$\frac{5}{4}$}\,x^2
\bar{\ell}_{\ind F}^{\;2}
+x^2 k_{\ind F}+O(x^3)\co\label{MF}\no
\bar{\ell}_{\ind M}\al=\al \frac{1}{51}\,(28\,\bar{\ell}_1+32\,\bar{\ell}_2-
9\,\bar{\ell}_3+49)\co\\
\bar{\ell}_{\ind F}\al=\al \frac{1}{30}\,(14\,\bar{\ell}_1+16\,\bar{\ell}_2+
6\,\bar{\ell}_3-6\,\bar{\ell}_4+23)\co
\nonumber\eea 
where
$M^2\!=\!2\, \hat{m}\, B$ is the leading 
term\footnote{Note that $B$ differs
  from the constant $B_0$ occurring  
in section \ref{pseudoscalar masses} by terms of $O(m_s)$.} in the 
expansion of $M_\pi^2$, while
$F$ is the pion decay constant in the limit
$\hat{m}=0$ at fixed $m_s$,   
and $x$ denotes the dimensionless ratio
\bea x =\left(\!\frac{M}{4\hspace{0.05em} \pi F}\!\right)^{\!\!2}\fs
\eea
In this representation, the chiral logarithms are hidden in the quantities
$\bar{\ell}_n$, which represent the running coupling constants at scale
$\mu=M$ and depend logarithmically on $M$
\bea\label{Lambda} \bar{\ell}_n\al=\al \ln \frac{\Lambda_n^2}{M^2}\fs\eea 
The mass-independent
terms $k_{\ind M}$ and $k_{\ind F}$ account for the remainder of $O(M^4)$, in
particular for the contributions from the effective couplings of 
${\cal  L}^{(6)}$.  

The relations (\ref{MF}) are dominated by the contributions of $O(M^2)$,
which involve only two parameters: the values of $\Lambda_3$ and $\Lambda_4$,
respectively. In effect, these two parameters replace the coefficients
$a_{\ind M},a_{\ind F}$ in the polynomial approximations $M_\pi^2=M^2(1+a_{\ind
  M} M^2)$,  
$F_\pi=F(1+a_{\ind F} M^2)$ that are sometimes used to perform the
extrapolation of lattice data. In contrast to these
approximations, the formulae (\ref{MF}) do account for the relevant infrared
singularities and are exact, up to and including $O(M^4)$.

\section{NUMERICAL DISCUSSION}
The estimates given in \cite{BCT} confirm the expectation that the
contributions from 
$k_{\ind M}$ and $k_{\ind F}$ are of order $(M/M_S)^4$, where $M_S\simeq 1
\,\mbox{GeV}$ is the mass scale characteristic of the scalar or pseudoscalar 
non-Goldstone states contributing to the relevant sum rules 
(in the SU(2) framework we are using here, the  
$K\bar{K}$ continuum also contributes to the effective coupling constants, but
in view of $4M_K^2\simeq M_S^2$, the relevant scale is even somewhat larger). 
Unless the
quark masses are taken much larger than in nature, these terms are very small
and can just as well be dropped. 
The relations (\ref{MF}) then specify $M_\pi$ and $F_\pi$ as functions of 
$\hat{m}$,
in terms of the 6 constants $F,B,\Lambda_1,\ldots\,,\Lambda_4$. 

Except for $B$ and $\Lambda_3$, all of these can be determined quite 
accurately from experiment: 
The scales $\Lambda_1$ and $\Lambda_2$ can be evaluated on
the basis of the Roy equations for $\pi\pi$ scattering, while $\Lambda_4$ 
is related to the scalar charge radius, which can be extracted from 
a dispersive analysis of the scalar form factor \cite{ACGL,CGL}. 
For the purpose of illustration, I use the lowest order relation 
$B\simeq (M_K^2-\frac{1}{2}M_\pi^2)/m_s$ to replace the variable 
$\hat{m}$ by the ratio $\hat{m}/m_s$. For $\Lambda_3$, I invoke the crude
estimate 
$\ln \Lambda_3^2/M_\pi^2=2.9\pm 2.4$ given in \cite{GL 1984}, which
amounts to $0.2\, \mbox{GeV}<\Lambda_3<2\,\mbox{GeV}$. 

Consider first the ratio $F_\pi/F$, for which the poorly known scale 
$\Lambda_3$ only enters at next-to-next-to leading order. 
The upper one of the two shaded regions in fig.~1 shows the behaviour of this
ratio as a function of $\hat{m}$, according to formula (\ref{MF}).
The change in $F_\pi$ occurring 
if $\hat{m}$ is increased from the physical value to $\frac{1}{2}\,m_s$ is of 
the expected size, comparable to the difference between $F_K$ and $F_\pi$.
The curvature makes it evident
that a linear extrapolation in $\hat{m}$ is meaningless. 
The essential para\-meter here is the
scale $\Lambda_4$ that determines the magnitude of the term of order $M^2$.
The corrections of order $M^4$ are small -- the scale relevant for these
is $\Lambda_F\simeq 0.5\,\mbox{GeV}$. 
\begin{figure}[t] 
\psfrag{y}{}
\psfrag{Fpi}{\hspace{-3em}\raisebox{-0.4em}{$F_\pi/F$}}
\psfrag{Mpi}{\hspace{-3em}\raisebox{-2.7em}{$
M_\pi^2/M^2$}}
\psfrag{m}{\raisebox{-1.3em}{\hspace{-10em}$\hat{m}/m_s$}}
\centering
\hspace*{0em}\mbox{\epsfysize=4.5cm \epsfbox{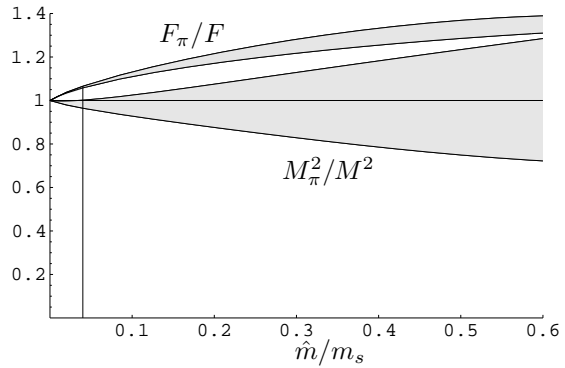} }

\vspace*{-1.5em}
\caption{Dependence of the ratios $F_\pi/F$ and $M_\pi^2/M^2$ on 
$\hat{m}=\frac{1}{2}(m_u+m_d)$. The strange quark mass is held fixed at the
physical value. The vertical line corresponds to the physical value of
$\hat{m}$.} 

\vspace*{-2.5em}
\end{figure}

In the case of the ratio $M_\pi^2/M^2$, on the other hand, the
dominating contribution is determined by the scale $\Lambda_3$. 
The fact that the information about that scale is very meagre
shows up through very large uncertainties. In particular, with 
$\Lambda_3\simeq 0.5 \,\mbox{GeV}$, the ratio $M_\pi^2/M^2$ would remain 
very close to 1, on the entire interval shown. The corrections of $O(M^4)$ are
small also in this case (the relevant scale is 
$\Lambda_M\simeq 0.6\,\mbox{GeV}$).

The above discussion shows
that brute force is not the only way to reach the very small values of $m_u$
and $m_d$ observed in nature. 
It suffices to equip the strange quark with the physical value of $m_s$ 
and to measure the dependence
of the pion mass on $m_u,m_d$ in the region where $M_\pi$ takes a value like
400 or 500 MeV. Since the dependence on the quark masses is known rather
accurately in terms of the two constants $B$ and $\Lambda_3$, a fit to the
data based on eq.~(\ref{MF}) should provide an extra\-po\-lation to the 
physical quark masses that is under good control. Moreover, the resulting 
value for $\Lambda_3$ would be of considerable interest, because that scale
also shows up in other contexts, in the $\pi\pi$ scattering lengths, 
for example. A measurement of the mass dependence 
of $F_\pi$ in the same region would be useful too, because
it would provide a check on the dispersive analysis of the scalar radius
that underlies the determiation of $\Lambda_4$ (in view of the strong
unitarity cut in the scalar form 
factor, a direct evaluation of the scalar radius on the lattice 
is likely more difficult).

\section{{\large\boldmath $\pi\pi$\unboldmath} SCATTERING LENGTHS}
As a further example, I consider the two $\pi\pi$ $S$-wave scattering lengths,
for which the dependence on the quark masses was discussed already in
\cite{Colangelo 1997}, including a comparison with the quenched lattice data
available at the time. 
The two loop representation \cite{BCEGS} explicitly specifies the scattering
lengths in terms of the effective coupling constants, up to and including
contributions of $O(M^6)$. 
Fig.~2 shows the corresponding
corrections to Weinberg's low energy theorem\footnote{The
standard definition of the scattering length 
corresponds to $a_0/M_\pi$. It is not suitable, because it
differs from the invariant 
scattering amplitude at threshold by a kinematic factor that diverges in the
chiral limit.}
\cite{Weinberg 1966}   
\bea I\!=\!0: a_0^{\mbox \tiny W}=\frac{7 M_\pi^2}{32 \,\pi \, F_\pi^2}\co
\hspace{1.3em}
I\!=\!2: a_0^{\mbox \tiny W}=-\frac{M_\pi^2}{16 \,\pi \,
  F_\pi^2}\fs\nonumber\eea 
Note that the Weinberg formulae are written in terms of the
values of $F_\pi$ and $M_\pi$ that correspond to the quark mass of interest --
instead of the corresponding lowest order terms $F$ and $M$. Although the
range shown here is considerably smaller than in
fig.~1, the corrections are much larger: The expansion of the
scattering lengths in powers of $M_\pi$ converges only very
slowly. 
The effect arises from the
chiral logarithms associated with the unitarity cut, which in the case of the
$I=0$ channel happen to pick up large coefficients. In fact, the 
chiral perturbation theory formulae underlying the figure are meaningful only 
in the range where the corrections are small (the shaded regions exclusively 
account for the 
uncertainties in the values of the coupling constants). 

I conclude that, in the case of the $I=0$ scattering length, the
extrapolation in $\hat{m}$ requires significantly smaller quark
masses than the one for $M_\pi$ or $F_\pi$. In the $I=2$ channel,
the effects are much smaller, because this channel is
exotic: The final state interaction is weak and repulsive.
\begin{figure}[t] 
\psfrag{I0}{\hspace{0.2em}\raisebox{1.3em}{$I=0$}}
\psfrag{I2}{\raisebox{0.7em}{$I=2$}}
\psfrag{a}{\large\hspace{-1.8em}\raisebox{-2em}{$\frac{\rule[-0.2em]{0em}{0em}
a_0}
{\rule{0em}{0.6em}a^{\mbox{\tiny W}}_0}$}}
\psfrag{m}{\hspace{-10em}\raisebox{-1.4em}{$M_\pi$ (MeV)}}

\vspace*{-1em}
\centering
\hspace*{1.4em}\mbox{\epsfysize=5cm \epsfbox{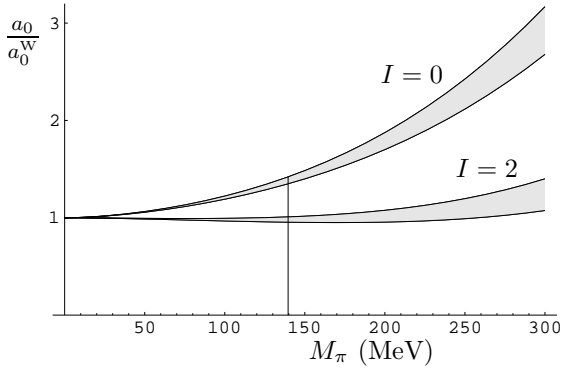} }
  
\vspace*{-1.5em}
\caption{$\pi\pi$ $S$-wave scattering lengths as a function of the pion mass.} 
\vspace*{-2em}
\end{figure}   

The method proposed in \cite{CGL} replaces the expansion of the
scattering amplitude at threshold by one in the unphysical region,  
where the convergence properties are similar to those for 
$F_\pi$ and $M_\pi^2$. Indeed, that method yields a remarkably precise 
prediction for the scattering lengths: $a_0^0=0.220\pm0.005$,
$a_0^2=-0.0444\pm 0.0010$. The lattice result, 
$a_0^2=-0.0374\pm 0.0049$ \cite{lattice
  I2}, is on the low side, but not
inconsistent with the prediction.

\section{CONCLUSIONS}
The results for $m_s$ obtained from recent lattice simulations with dynamical
quarks are consistent with those based on sum rules, but the central 
values are on the low side. 

The lattice results for the ratios $m_u/m_d$ and $m_s/m_d$ confirm
the values obtained in the framework of chiral perturbation theory.

The dependence of the various observables on the quark masses contains
   infrared singularities which can be worked out by means of chiral
   perturbation theory. This information might be useful for the extrapolation
   of lattice results to the small quark mass values of physical interest.\\

It is a pleasure to thank B.~Ananthanarayan,
   J.~Pasupathy, A.~Patel and Ch.~Shukre for their 
hospitality during my stay at the
Indian Institute of Science, G.~Colangelo, J.~Gasser and S.~Narsion for
useful comments and the Swiss 
National Foundation for support.


\begin{thebibliography}{99}
\bibitem{Phys Rep}
The literature contains even earlier estimates, which however were based on the
assumption that the strong interaction conserves isospin and thus took
$m_u\!=\!m_d$ for granted.  The prehistory is reviewed in
J.~Gasser and H.~Leutwyler, Phys.\ Rep.\ 87 (1982) 77.

\bibitem{GL75}
J.~Gasser and H.~Leutwyler, Nucl.\ Phys.\ B94 (1975) 269.

\bibitem{Weinberg1977}
S.~Weinberg, in {\it A Festschrift for I.\ I.\ Rabi}, ed.
L.\  Motz, Trans.\ New York Acad.\ Sci.\ Ser.\ II 38 (1977) 185.

\bibitem{Manohar}A.~Manohar, in {\it Review of Particle Physics}, 
Eur.\ Phys.\ J.\ C15 (2000) 377.

\bibitem{Descotes Stern}
S.~Descotes and J.~Stern,
Phys.\ Lett.\  B488 (2000) 274

\bibitem{QCD SR1} A.~I.~Vainshtein et al., Sov.\ J.\ Nucl.\ Phys.\ 27
(1978) 274; 
B.~L.~Ioffe, Nucl.\ Phys.\ B188 (1981) 317;  B191 (1981) 591(E).

\bibitem{CK}
P.~Colangelo and A.~Khodjamirian,
hep-ph/0010175.

\bibitem{Pich Prades}
A.~Pich and J.~Prades,
JHEP 9910 (1999) 004.

\bibitem{Kambor Maltman}J.~Kambor and K.~Maltman,
Phys.\ Rev.\  D62 (2000) 093023.

\bibitem{ALEPH}R.~Barate et al.~[ALEPH Collaboration],
Eur.\ Phys.\ J.\ C11 (1999) 599. 


\bibitem{Narison}S.~Narison,
Nucl.\ Phys.\ Proc.\ Suppl.\  86 (2000) 242.

\bibitem{ms sum rules}(only references since 1995)
M.~Jamin and M.~M\"unz,
Z.\ Phys.\  C66 (1995) 633;
S.~Narison,
Phys.\ Lett.\ B358 (1995) 113; 
B466 (1999) 345;
J.~Bijnens, J.~Prades and E.~de Rafael,
Phys.\ Lett.\  B348 (1995) 226;
K.~G.~Chetyrkin, C.~A.~Dominguez, D.~Pirjol and K.~Schilcher,
Phys.\ Rev.\  D51 (1995) 5090;
K.~G.~Chetyrkin, D.~Pirjol and K.~Schilcher,
Phys.\ Lett.\  B404 (1997) 337;
P.~Colangelo, F.~De Fazio, G.~Nardulli and N.~Paver,
Phys.\ Lett.\  B408 (1997) 340;
T.~Bhattacharya, R.~Gupta and K.~Maltman,
Phys.\ Rev.\  D57 (1998) 5455;
M.~Jamin,
Nucl.\ Phys.\ Proc.\ Suppl.\  64 (1998) 250;
K.~G.~Chetyrkin, J.~H.~K\"uhn and A.~A.~Pivovarov,
Nucl.\ Phys.\  B533 (1998) 473;
K.~Maltman,
Phys.\ Lett.\ B428 (1998) 179; B440 (1998) 367; B462 (1999) 14; 
B462 (1999) 195; Phys.\ Rev.\ D58 (1998) 093015.

\bibitem{lower bound}L.~Lellouch, E.~de Rafael and J.~Taron, 
Phys.\ Lett.\ B414
  (1997) 195;
F.~J.~Yndurain, Nucl.\ Phys.\ B517 (1998) 324;
H.~G.~Dosch and S.~Narison, Phys.~Lett.~B417 (1998) 173.

\bibitem{Dashen}R.~Dashen, Phys.\ Rev.\ 183 (1969) 1245.

\bibitem{Weinberg Physica}
S.~Weinberg, Physica A96 (1979) 327.

\bibitem{GL SU(3)}J.~Gasser and H.~Leutwyler,
Nucl.\ Phys.\ B250 (1985) 465.

\bibitem{Kaplan Manohar}D.~B.~Kaplan and A.~V.~Manohar, Phys.\ Rev.\ Lett.\ 56 
(1986) 2004.

\bibitem{Light quark masses}H.~Leutwyler, Phys.\ Lett.\ B378 (1996) 313.

\bibitem{GL eta}J.~Gasser and H.~Leutwyler,
Nucl.\ Phys.\ B250 (1985) 539.

\bibitem{Baur Kambor Wyler}
R.~Baur, J.~Kambor and D.~Wyler,
Nucl.\ Phys.\  B460 (1996) 127.


\bibitem{Kambor Wiesendanger Wyler}
J.~Kambor, C.~Wiesendanger and D.~Wyler,
Nucl.\ Phys.\  B465 (1996) 215.

\bibitem{AL}
A.~V.~Anisovich and H.~Leutwyler, Phys.\ Lett.\ B375 (1996) 335. 

\bibitem{Banks Nir Seiberg} For a review, see T.~Banks, Y.~Nir and 
N.~Sei\-berg, in {\it Yukawa couplings and the origin of
mass}, ed. P. Ramond
(Int. Press, Cambridge MA, 1995), hep-ph/9403203.
 
\bibitem{Kaiser Leutwyler}R.~Kaiser and H.~Leutwyler, 
hep-ph/0007101 
and in {\it Nonperturbative Methods in Quantum Field
Theory}, 
eds. A.W.\ Schreiber, A.G.\ Williams and A.W.\ Thomas (World Scientific, 
Singapore, 1998), hep-ph/9806336.

\bibitem{Montpellier}H.~Leutwyler,
Nucl.\ Phys.\ Proc.\ Suppl.\  64 (1998) 223.

\bibitem{GL 1984}J.~Gasser and H.~Leutwyler, Annals Phys.\  158 (1984) 
142.

\bibitem{Ecker}G.~Ecker et al., Nucl.\  Phys.\  B321 (1989) 311; Phys.\ 
Lett.\  B223 (1989) 425.

\bibitem{Leutwyler1990}
H.~Leutwyler, Nucl.\ Phys.\ B337 (1990)
108.

\bibitem{ABT}G.~Amor\'os, J.~Bijnens and P.~Talavera,
Nucl.\ Phys.\  B568 (2000) 319;
B585 (2000) 293;
Phys.\ Lett.\  B480 (2000) 71.
Among other things, these authors also discuss 
the dependence of various quantities on $m_s$ (compare sections 9 and 10). 

\bibitem{finite volume} J.\ Gasser and H.\ Leutwyler, Phys.\ Lett.\ B84 (1987)
  83; B188 (1987) 477.

\bibitem{Buergi}U.~B\"urgi,
Nucl.\ Phys.\  B479 (1996) 392.

\bibitem{BCEGS}
J.\ Bijnens, G.\ Colangelo, J.\ Gasser and M.~E.\ Sainio,
Phys.\ Lett.\  B374 (1996) 210,
Nucl.\ Phys.\   B508 (1997) 263;
B517 (1998) 639 (E). 

\bibitem{BCE}J.~Bijnens, G.~Colangelo and G.~Ecker, JHEP 9902 (1999) 020;
Annals Phys.\ 290 (2000) 100.
 
\bibitem{BCT} J.~Bijnens, G.~Colangelo and P.~Talavera,
JHEP 9805 (1998) 014.

\bibitem{ACGL}
B.~Ananthanarayan, G.~Colangelo, J.~Gasser and H.~Leutwyler,
hep-ph/0005297.

\bibitem{CGL}G.~Colangelo, J.~Gasser and H.~Leutwyler,
Phys.\ Lett.\  B488 (2000) 261. A more detailed account of this work,
in particular also of the results obtained for the effective coupling
constants, is in preparation.

\bibitem{Colangelo 1997}G.~Colangelo, Phys.\ Lett.\ B395
(1997) 289.

\bibitem{Weinberg 1966}
S.~Weinberg,
Phys.\ Rev.\ Lett.\  17 (1966) 616.

\bibitem{lattice I2}
S.~Aoki {\it et al.}  [JLQCD Collaboration],
Nucl.\ Phys.\ Proc.\ Suppl.\  83 (2000) 241.


\end{thebibliography}
\end{document}